# A Diabatic Three-State Representation of Photoisomerization in the Green Fluorescent Protein Chromophore

*Seth Olsen and Ross H. McKenzie*

The University of Queensland, Centre for Organic Photonics and Electronics and School of Mathematics and Physics, Brisbane, QLD 4072 Australia

s.olsen1@uq.edu.au

**Abstract** We give a quantum chemical description of the photoisomerization reaction of green fluorescent protein (GFP) chromophores using a representation over three diabatic states. Photoisomerization leads to non-radiative decay, and competes with fluorescence in these systems. In the protein, this pathway is suppressed, leading to fluorescence. Understanding the electronic states relevant to photoisomerization is a prerequisite to understanding how the protein suppresses it, and preserves the emitting state of the chromophore. We present a solution to the state-averaged complete active space problem, which is spanned at convergence by three fragment-localized orbitals. We generate the diabatic-state representation by block diagonalization transformation of the Hamiltonian calculated for the anionic chromophore model HBDI with multi-reference, multi-state perturbation theory. The diabatic states are charge-localized and admit a natural valence-bond interpretation. At planar geometries, the diabatic picture of the optical excitation reduces to the canonical two-state charge transfer resonance of the anion. Extension to a three-state model is necessary to describe decay via *two* possible pathways associated with photoisomerization of the (methine) bridge. Parametric Hamiltonians based on the three-state ansatz can be fit directly to data generated using the underlying active space. We provide an illustrative example of such a parametric Hamiltonian.

## 1. Introduction

The discovery and development of the green fluorescent protein (GFP)[1] and its homologues[2] has revolutionized biotechnology, cell biology and molecular biophysics. The proteins are useful because they become fluorescent automatically following expression and folding. The chromophore is usually derived from a common *p*-hydroxybenzylidene-imidazolinone (HBI) motif[2], which, unmodified, is the chromophore of the most common (possibly ancestral[3]) green subclass. This motif can occur in multiple titration states[4]. In green fluorescent proteins (GFPs), the emitting state has been assigned to an anionic HBI species[5]. In some variants, this species is produced in the excited state following excitation of a ground neutral-like[4] form in a manner consistent with excited-state proton transfer[6]. Modelling and and understanding the photochemistry and photophysics of complex molecular materials like proteins is a major challenge to theory[7,8].

The utility of fluorescent proteins is intimately related to their emission. Understanding the nature of the emitting state is a major goal for theoretical models[9]. A central question is how the emitting state is generated and preserved, because synthetic models such as dimethyl-HBI (HBDI, figure 1)[4] and denatured proteins[10] do not fluoresce under normal conditions[4,11]. Spectral similarity suggests that the emitting state is localized to the chromophore[4,11], so this question can be further broken down into three sub-questions: *What is the state of the chromophore in the protein emitting state? Why is this state not observed in chromophores outside of the protein environment? How does the protein generate and preserve this state so that high-yield emission is achieved?*

A bright optical excitation is common in methine dyes[12] such as HBI anion, and is related to a charge transfer resonance[13], shown in figure 1. As in other fluorogenic monomethine dyes[14,15], a significant non-radiative decay mechanism is cis-trans photoisomerization of the bridge[16]. This mechanism leads to ultrafast internal conversion of model compounds in solution[17,18]. Crystal structures of proteins with reduced emission or detectable non-emitting states show a distribution of isomeric states of the chromohore[19-22], suggesting that this non-radiative decay channel can also operate in the protein. Accordingly, it is normally considered that the protein preserves the emitting state by restricting the ability of the chromophore to decay via photoisomerization. This idea supported by analysis of the fluorescence structure of protein and chromophore at low temperatures[11], and predicted emission energy of chromophore models that are relaxed on the excited state under constraint of planarity[9,23].

The physical mechanism by which the protein constrains the chromophore is still not understood. Compounding the mystery is the fact that torsion about *either* bridge bond[9,24,25], or possibly a combination[9], can quench the emitting state and lead to nonradiative decay. Presumably, the protein must inhibit displacement along *all* bridge torsion coordinates in order to preserve the emitting state. This "multiple pathway problem" has been highlighted by Zimmer and coworkers, whose molecular mechanics model simulations suggest steric hindrance by the protein is not complete[26-29]. However, these models cannot address the twisted intramolecular charge-transfer (TICT) character that accompanies excited-state torsion[23,24]. Straightforward extension of theories of enzymatic catalysis[30] would suggest that electrostatic interactions between the TICT states and the protein may contribute to suppression or control of the photoisomerization. Similar physics has been invoked to explain fast decay of GFP chromophores in solution simulations[31]. More recent simulations suggest that this mechanism may control photoisomerization in photoactive yellow protein chromophores[32].

In order to address the importance of charge-transfer for preservation of the emitting state, it is necessary to have models that can represent these effects. This is, in principle, possible with 'quantum mechanical/molecular mechanical' (QM/MM) methods[33], but these techniques (as generally applied) are so expensive as to prohibit sufficient sampling[34]. This is a serious drawback, as undersampling can lead to misleading or incomplete results[35]. In specific case of fluorescent proteins, the need for sampling is implied by blinking observed in single molecule experiments[36-38] and the discretization of transition rates between states with different emission colors[39]. Recent work has shown that extensive sampling is required to reproduce intra-protein electrostatic fields measured by vibrational Stark spectroscopy[40].

Methods based upon parametric electronic Hamiltonians offer a route to models that can describe electronic state changes without sacrificing sampling accuracy. A good example is the empirical valence bond (EVB) method[41]. Other examples include Heisenberg spin models[42]. These methods define a Hamiltonian over a suitably chosen space of valence-bond wavefunctions. The electrostatic effects of the environment on the electronic structure are represented by differential stabilization of covalent and ionic states. The Hamiltonian matrix elements can be expressed using parameterized analytical functions of the nuclear coordinates, making the method efficient. It can easily be adapted to the functional forms used in common molecular mechanics fields without too much loss of accuracy[43]. A primary step in the formulation of an EVB model is the inference of a suitable set of valence states. Such inference relies ultimately on the researcher's intuition, but such intuition can, and should, be informed by more objective sources, such as solutions from *ab initio* electronic structure calculations. Unfortunately, such calculations do not always lend themselves easily to conceptual interpretations. Therefore, when such conceptual models *do* emerge, it is of immediate interest to the field, and a catalyst for further development.

In this paper, we extract a simple diabatic representation of the photoisomerization reaction from *ab initio* electronic structure calculations on the ground and excited states of a GFP chromophore model. The representation emerges from a block-diagonalization[44] of the electronic Hamiltonian for the HBDI anion, calculated by multi-state *ab initio* calculations, which include both static and dynamic correlation effects. The states in the diabatic ansatz for HBDI can be mapped onto states that arise in the valence-bond resonance of allylic anions[45]. It has properties that are useful for the development of conceptual and parametric electronic structure models. In particular, the states generated by the transformation are chemically localized, facilitating the development of models parameterized by the molecular geometry.

In Section 2, we describe computational electronic structure calculations, which we used to generate a diabatic representation of the photoisomerization of HBI anion. In Section 3, we describe the block diagonalization transformation used to generate the diabatic representation from the electronic structure results. In Section 4 we describe the diabatic states that emerge from the transformation. In sections 5 and 6 we discuss the structure of the eigenvectors of the Hamiltonian in the diabatic representation, and the structure of the Hamiltonian matrix in this representation. In section 7 we will discuss implications for model development, and provide a simple example of how parametric models may be generated.

Section 8 discusses the ansatz in relation to other models of photoisomerization. Section 9 outlines the limits of applicability of the three-state ansatz. We conclude in section 10.

## 2. Quantum Chemistry Calculations

This paper describes a diabatic representation of a solution to a state-averaged[46,47] complete active space self-consistent field)[48,49] (SA-CASSCF) problem for HBDI anion. The four electron, three orbital active space is summarized in figure 2. It is isomorphic to the π system of an allyl anion. The CASSCF optimized for an average over three states(SA3-CAS(4,3)). Adiabatic energies and state-specific properties obtained via identical methodology have been described earlier[23]. This paper is an account of a diabatic representation of the same electronic structure, which was not previously described. At convergence, the active space is spanned by Foster-Boys localized[50] orbitals on the phenoxy, methine bridge, and imidazolinone moieties. This structure is adapted to the charge transfer resonance (Figure 1), which is a distinguishing feature of monomethine dyes[12]. To the SA-CASSCF solution, we applied multi-state, multi-reference perturbation theory[51] (MR-MS-RSPT2), in which the highest-lying 32 occupied orbitals were correlated. A Dunning DZP basis[52,53] was used. All of the electronic structure calculations in question were performed with Molpro[54]. Geometries, SA-CASSCF and MR-MS-RSPT2 energies, and state-averaged natural orbitals and occupation numbers are outlined in the Supplement[55].

The Boys-localized active spaces are isomorphic over a broad distribution of bridge twisted structures (Figure 3). The self consistent field is nonlinear, and multiple solutions can be found; state-averaged natural orbitals and occupation numbers, sufficient to specify the wavefunction, in available in the Supplement[55]. All calculations were performed in the $C_1$ point group (i.e. no symmetry was used).

The calculations were performed at a handful of relevant geometries optimized on the SA3-CAS(4,3) surfaces. We will refer to these using the same nomenclature as in Reference 23. The geometries included the ground state minima of the Z and E isomers (Z- and E-Min-$S_0$), $S_1$ minima twisted about the phenoxy-bridge (Z,E-P-$S_1$) and imidazolinone-bridge bond (I-$S_1$), and a structure which was generated via minimization on $S_1$ under the constraint of disrotatory (hula) twist of both bonds by $90^0$ (HT-$S_1$). In addition, a variety of other geometries were optimized on the SA-CASSCF $S_1$ surface with constraint on the bridge bond torsion angles. Constrained optimizations were also carried out at points where the two bridge torsion angles ($\theta_I$ and $\theta_P$, the torsions about the phenoxy-bridge and imidazolinone-bridge bonds) were set to values at $(0^0,0^0)$, $(0^0,180^0)$, $(0^0, 45^0)$, $(0^0,135^0)$, $(180^0,45^0)$, $(180^0,135^0)$, $(45^0, -45^0)$, $(135^0,-135^0)$, $(45^0,45^0)$, $(135^0,135^0)$, and $(90^0, -90^0)$. The MR-MS-RSPT2 results at these points were used as data to fit a parametric surface for the photoisomerization, which we describe at the end of this paper.

Over the fragment-localized orbitals that span the active space at convergence, we define singlet configuration state functions (CSFs) as in Figure 2. The space of CSFs is spanned by six states generated by permuting four electrons in the three localized orbitals in manners consistent with singlet spin. They can be categorized into 'covalent' configurations and 'ionic' configurations. In the covalent configurations, one of the three orbitals is doubly filled while the other two are singly occupied and coupled to form a singlet. In the ionic configurations, two of the orbitals are doubly filled and one is empty. This is precisely the set that would be generated by considering CSFs over the carbogenic p orbitals of an allyl anion. The covalent CSFs carry a net charge, and the ionic configurations arise from polarizing the singlet pairs in the covalent CSFs. It is important to emphasize that each ionic configuration can be generated in this way from *two different covalent CSFs*. This relationship between ionic and covalent CSFs is emphasized in figure 2 with double-headed arrows. Note that the covalent CSFs do not carry any bond stabilization energy without interaction with the ionic CSFs. This is always true for covalent valence-bond states defined over *orthogonal* orbitals. In this type of ansatz (unlike traditional *non*-orthogonal valence bond ansatzes), bonding stabilization is expressed by superpositions of covalent configurations and ionic configurations that polarize the bond. For an accessible discussion of this point see Reference [56]. A more advanced treatment can be found in Reference [57].

The optical excitations of HBDI[23,58,59] and HBI[9,31,60,61] anion (and other FP chromophore anions) have been studied using multi-reference perturbation theory previously, using larger[9,60] and smaller[31,59] active spaces than in Figure 2. In our experience[62], any such solution targeting the ground and lowest π-π* states will have significant projection on an active space analogous to Figure 2. The gas phase absorption of HBDI anion has a maximum at 2.59 eV (479 nm) with a 45 nm full width at half max (2.85 eV and 2.41 eV)[63]. Table 1 lists a collection of published estimates of the $S_0$-$S_1$ energy splitting of HBDI and HBI anion, obtained with different SA-CASSCF and MR-RSPT2 methods. There is no clear correlation between accuracy and the dimension of the active space. The operators defined by a CAS expansion span a Lie algebra[64], and the self-consistent field can be interpreted as maximum entropy inference subject to constraint[65]. Larger active spaces represent more constraints than small ones. Nemukhin has shown, for a kindling fluorescent protein (KFP) chromophore anion, that the excitation energy is insensitive to the active space dimension[66]. We have not been able to obtain, with a smaller active space, a diabatic representation that is continuous with respect to twisting of both bridge bonds on the $S_1$ surface. Qualitatively, the active space in Figure 2 seems to contain not much more information than implied by the resonance in Figure 1. Charge transfer resonances such as in Figure 1 are at the foundation of our understanding of the electronic properties of conjugated dyes[13].

It is known that calculations of the ππ* excited states of ethylene (a model for other conjugated systems) are very sensitive to the quality of the basis set used, due to the very different spatial extent of these states relative to the ground state and low-lying Rydberg states[67]. A larger basis set would provide more accurate approximation of the exact Born-Oppenheimer states for HBDI anion.

### 3. The Block-Diagonalization Transformation

We have generated the diabatic basis with a unitary block-diagonalization transformation, which has been discussed by Cederbaum et al.[44] and by Pacher et al.[68]. We apply the block transformation to the CSF basis over Boys localized orbitals, with localization performed independently at different geometries. The localization is performed separately within each of the invariant spaces of the CASSCF wavefunction. A similar but more sophisticated approach, based on overlap of solutions at neighboring geometries, has been discussed and applied by Domcke et al[69]. The orbitals obtained via the Boys procedure are similar at different geometries, and are sufficient to demonstrate the representation in this case. The domain of the transformation is the space of CSFs, and the block diagonalization is performed between the ionic and covalent CSF subspaces (see Fig. 2). The transformation is unitary, and works as follows. Consider a matrix **C**, whose columns are the energy eigenstates in the CSF basis, and the diagonal matrix **V**, whose entries are the eigenenergies of the system. We block-diagonalize the six-by-six Hamiltonian in the CSF representation into two three-by-three blocks. If we label the blocks by the letters α and β, and use the notation $\mathbf{C}_{\alpha\alpha}$ to refer to the α block of **C**, $\mathbf{C}_{\beta\beta}$ to refer to the β block of **C**, $\mathbf{C}_{\alpha\beta}$ to refer to the α-β off-diagonal block, etc. Then the lower block $\mathbf{H}_{\alpha\alpha}$ of the block-diagonalized Hamiltonian defines an effective Hamiltonian $\mathbf{H}_{eff}$, which is written as (1).

$$\mathbf{H}_{eff} \equiv \mathbf{H}_{\alpha\alpha} = \left(\mathbf{C}_{\alpha\alpha}\mathbf{C}_{\alpha\alpha}^{\dagger}\right)^{1/2}\left(\mathbf{C}_{\alpha\alpha}^{-1}\right)^{\dagger}\mathbf{V}_{\alpha\alpha}\mathbf{C}_{\alpha\alpha}^{-1}\left(\mathbf{C}_{\alpha\alpha}\mathbf{C}_{\alpha\alpha}^{\dagger}\right)^{1/2} \quad (1)$$

There is an analogous expression for the block $\mathbf{H}_{\beta\beta}$, if desired. The diabatic states are given by the columns of the transformation matrix (2).

$$\mathbf{R}_{\alpha\alpha} = \mathbf{C}_{\alpha\alpha}^{-1}\left(\mathbf{C}_{\alpha\alpha}\mathbf{C}_{\alpha\alpha}^{\dagger}\right)^{1/2} \quad (2)$$

This is the transformation that does as little else as possible other than block-diagonalize the Hamiltonian in the CSF basis[44]. The transformation is unitary. Non-unitary versions have also been discussed, but what is particularly useful about the unitary form is that it can be defined using only information pertaining to the energies and eigenstates of the lower block (as expressed here). This is not true of the non-unitary transformation[44].

Here, we have used the eigenenergies of the MR-MS-RSPT2 calculation to generate **V** and have used the 'eigenstates' of the MR-MS-RSPT2, given by the reference SA-CASSCF eigenstates multiplied by the MR-MS-RSPT2 mixing matrix, to generate **C**. We used an approximation wherein the perturbation theory was applied using the canonical orbitals, and the mixing of the reference eigenstates thus calculated was applied to their representation in the localized orbital basis.

As argued by Pacher et al., this transformation produces states that are approximately diabatic in nature[68]. The diabaticity must be approximate, because the nonadiabatic coupling vector field has a component that cannot be removed by transformation in a subspace of electronic states[70]. We refer to the representation as 'diabatic' for simplicity here.

## 4. Structure of the Diabatic States

The electronic structure of the diabatic states, which emerge from the block diagonalization, is listed in Tables 2-4 at a collection of representative geometries of HBDI. Some of this data is also displayed visually in Figure 3, which also displays the underlying Boys localized orbitals at representative geometries with significant bridge twist. Each of the three diabatic states is dominated by a single covalent configuration in the localized orbital basis. This is true regardless of whether the block-diagonalization is performed at a planar ($Z,E$-Min-$S_0$) or twisted geometry ($Z,E$-P-$S_1$, I-$S_1$). This allows us to identify each of the diabatic states with the doubly filled orbital in the dominant covalent CSF for that state. We will label the diabatic states dominated by $|ppbi\rangle$, $|bbpi\rangle$, and $|iipb\rangle$ as $|P\rangle$, $|B\rangle$ and $|I\rangle$, respectively.

The diabatic states change their structure somewhat over the relevant set of configurations. These changes are continuous, and straightforward to describe and understand. At planar geometries, there are larger contributions from the ionic configuration state functions (CSFs). For the states $|P\rangle$ and $|I\rangle$, the ionic CSFs with the largest contribution are those associated with polarization of the singlet pair in the dominant covalent CSFs ($|ppbi\rangle$ and $|iipb\rangle$, respectively). This indicates that the singlet pairs form true chemical bonds at these geometries. Ionic CSFs also contribute to the $|B\rangle$ state at planar geometries. However, in the $|B\rangle$ state, the ionic CSFs contributing most are not those that polarize the singlet pair in the dominant covalent CSF ($|bbpi\rangle$). Instead, they are associated with polarization transverse to the bridge, or a simultaneous transfer of the electrons on the methine bridge to each of the rings. This indicates the absence of a true chemical bond in the $|B\rangle$ state at planar geometries. This is another way of saying that $|B\rangle$ is a biradical[71].

At twisted geometries, the contribution from ionic CSFs which polarize the twisted bond are dramatically reduced. The diabatic states that feature singlet pairing across the twisted bond lose almost all of their projection on the ionic CSFs. At the twisted $S_1$ minima, they are completely dominated by a single covalent CSF. This indicates that these states represent biradical electronic structure across the twisted bond at these geometries. The diabatic state whose singlet pair does *not* cross the twisted bond maintains its projection on the ionic CSF space. The ionic CSFs that contribute are those associated with polarization of the singlet pair, so there is a chemical bond associated with this diabatic state.

Assignment of the diabatic states to Lewis structures is straightforward, and is outlined in Figure 4 for a planar case – the ground state minimum of the Z isomer (Z-Min-$S_0$), and a twisted case – an imidazolinone-twisted $S_1$ minimum (I-$S_1$). The position of the formal charge is dictated by the identity of the doubly filled fragment orbital, which is preserved across the relevant geometries. The nature of the singlet pair as a covalent bond or biradical structure is indicated by the presence or absence of appropriate ionic contributions. The pair can be represented as either line or a pair of dots, respectively, in the usual way.

The Lewis structural interpretation recaptures the canonical resonance of the HBDI anion (figure 1) at planar geometries. The $|P\rangle$ and $|I\rangle$ states are to be identified with the Lewis structures for the resonance as it is normally written (fig 1). The $|B\rangle$ state can be associated to a negatively charged bridge and a biradical interaction between the rings. An analogous state arises in resonance theories of allyl ions[45].

**5. Energy Eigenstates in the Diabatic Representation**

The electronic structure of the adiabatic states in the diabatic block diagonal state basis is summarized in Tables 5-7 for the same set of representative geometries as used in Tables 2-4. They can be categorized by whether the geometry is planar or twisted.

At planar geometries, the lowest two energy eigenstates ($S_0$ and $S_1$) are superpositions of the $|P\rangle$ and $|I\rangle$ diabatic states, while the third is dominated by the $|B\rangle$ diabatic state. If one assigns Lewis structures to the diabatic states as outlined above and in Figure 4, the ground state at these geometries can be identified with the resonance usually drawn for HBDI anion. The $S_1$ state is the 'twin state', formed by switching the parity of the superposition between $|P\rangle$ and $|I\rangle$[72]. Comparing the Z and E-Min-$S_0$ geometries highlights the diabatic nature of the block-diagonalized states. The structure of the $S_0$ state of the Z isomer is diabatically connected to the $S_1$ state of the E isomer and vice versa. The $S_2$ adiabatic

state is dominated by the $|B\rangle$ state at planar geometries, with only small contributions from the other diabatic states.

The picture that emerges is outlined in Figure 5. The ground state minimum of the Z isomer (Z-Min-$S_0$) and the imidazolinone twisted biradicaloid $S_1$ minimum (I-$S_1$) are used as examples to illustrate the structure of the states at planar and twisted geometries respectively. This highlights the 'twin state' nature of the $S_0$ and $S_1$ states at planar geometries and the $S_1$ and $S_2$ states at planar and twisted geometries, respectively. The 'twin state' concept has been used by Shaik and coworkers to explain the presence of elevated vibrational frequencies in excited state of aromatics[73-75] and linear polyenes[76] and by Zilberg and Haas[72,77,78] in the prediction of locations of conical intersections. The idea is that for a resonant ground state of a given parity, the anti-resonant (parity-reversed) superposition will describe a low-lying excited state. Similar results for other planar and twisted geometries can be drawn by consultation of tables 4-6.

**6. The Lower-Block Effective Hamiltonian**

The effective Hamiltonian produced by the diabatic transformation is summarized in table 8 at the same set of representative geometries used in tables 2-7. Again, the data is most easily summarized by categorization into planar and twisted geometries. Twisting a bond by $90^0$ decouples one of the diabatic states from the other two. The decoupled state is that which places the formal charge on the twisted fragment. Conversely, the coupling of the remaining two states becomes stronger. At the biradical $S_1$ minima its magnitude is on the same scale as the diagonal elements. As the twist angle increases past $90^0$, the coupling to the twisted fragment returns with opposite sign.

When the molecule is planar, the $|I\rangle$ and $|P\rangle$ diabatic states are nearly degenerate. When one of the bridge bonds twists, this degeneracy is broken. The splitting of the $|P\rangle$ and $|I\rangle$ states depends on the twist distribution of the bridge. Of this pair, the structure that ends up lowest in energy is the one without a singlet pair across the twisted bond. This implies that the $|I\rangle$ and $|P\rangle$ switch their energetic ordering in the vicinity of planar configurations. Our experience is that this crossing is avoided at energetically accessible planar configurations. As can be seen in table 8, the $|P\rangle$ and $|I\rangle$ states are also nearly degenerate at geometries with strong disrotatory (hula) twist. However, because the states at these geometries are not stabilized by interaction with the ionic CSFs (see, for example, Figure 3), their energy is higher at these geometries than when the molecule is planar.

### 7. Implications for Model Development

We have applied the Pacher-Cederbaum[44,68] block diagonalization transformation to a solution[23] of the state-averaged four-electron-in-three-orbital SA-CASSCF problem for a model of the green fluorescent protein chromophore. By doing so, we have managed to extracted a simple picture of the electronic structure of the ground and two lowest singlet excited states in terms of three diabatic states with charge-localized character. Two of these diabatic states are recognizable as the 'canonical resonance structures' of the green fluorescent protein chromophore anion. Therefore, we have established that a particular solution to the SA-CASSCF problem exists which corresponds to the traditional resonance picture. We did not impose the canonical resonance picture into the calculations. We have only used the inherent freedom to perform unitary transformations within the different orbital spaces. The resonance picture is naturally contained within the structure of the SA-CASSCF solution.

The localized valence-bond-like nature of the diabatic picture suggests that parametric approximations to the *ab initio* effective Hamiltonian may be applied to study the photoisomerization process in complex environments such as proteins, solutions and glasses. Because the underlying orbital set is localized in space and can be identified with localized molecular fragments, the Hamiltonian matrix elements can be expressed as parameterized functions of the molecular geometry. We anticipate that the ansatz can be applied in the context of a parametric representation of the electronic Hamiltonian as in the EVB method[41], or interfaced to time-dependent solvent models as has recently been done for models of photoisomerization in non-methine-bridged systems[79-81]. Such parametric Hamiltonians may be interfaced to classical or quantum mechanical models of the nuclear motion to model the photochemistry beyond the Born-Oppenheimer approximation.

To illustrate how the ansatz described here can be used to generate simplified models of the photoisomerization reaction, we have fitted a parametric model Hamiltonian via least-squares fitting to a set of lower-block Hamiltonians calculated at a collection of geometries. The geometries were relaxed on the $S_1$ surface under constraint of the bridge torsion. The functional forms of the Hamiltonian elements were polynomials in trigonometric functions of the bridge torsion angles, which were indicated by second-order quasi-degenerate perturbation[82] of the covalent block by the ionic block. The data used to fit the surfaces, as well as functional forms and details of the fitting procedure, are detailed in the Supplement[55]. Our purpose here is not to compare different possible functional forms for model Hamiltonians, nor different ways of performing the parameterization, but rather to establish that *simple* parameterizations can describe *gross* phenomena such as the multiplicity of pathways or the charge-localization upon twisting.

The adiabatic surfaces generated by the fitted Hamiltonian are shown in Figure 6. Also shown are the densities of the diabatic states in $S_1$ over the domain of torsion of both the bridge bonds. As can be seen in the figure, even this simple fit model can capture the gross features of the potential surfaces, such as the existence of two favorable pathways and the localization of formal charge in twisted regions of configuration space.

**8. Relationship to Other Models**

The photoisomerization of a *single* double bond, and chemistries involving a single biradicaloid structure, can be qualitatively described within a two electron, two orbital (2,2) model, as has been described in a beautiful paper by Michl and coworkers[71]. The (4,3) model space that we describe here corresponds to two such spaces sharing an orbital. This has important consequences. In the (2,2) model, a rigid torsion of a homopolar double bond does not lead to charge separation without an asymmetric perturbation (such as one-sided pyramidalization in ethylene and stilbene[83]). In our model, twisting leads to charge-transfer even in the absence of external perturbations, and even if the sites are identical, and the polarity of charge-transfer depends on which bond twists.

In a two-state model, the conditions for conical intersection require zeroing both the diagonal splitting and the off-diagonal coupling[71,84]. A three-state traceless Hamiltonian has five degrees of freedom[85]. The branching space of a three-state intersection has five dimensions[86] and spans the branching spaces of two intersecting two-state seams[87]. At twisted geometries, the Hamiltonian in our model is approximately block diagonal with a two-dimensional block ($S_1$/$S_2$) and a one-dimensional block ($S_0$). Approximately, this means that the adiabatic states can be obtained by diagonalization of the upper block alone. A conical intersection will arise when the energy lowering of $S_1$ relative to the middle diabatic state equals the diagonal splitting between the lowest two diabatic states.

For a constant coupling within the upper block, $S_0$/$S_1$ degeneracy can be achieved at twisted geometries by lowering the mean energy of the upper block or by raising the energy of the lowest diabatic state. The blocks differ by charge transfer across the twisted bond, so there is scope for induction of an intersection by an external field. As the charge transfer depends on which bond twists, the field that induces an intersection for one bond may not do so for the other. This is also true of an *internal* field, such as provided by substituent effects[88]. This latter point explains some of the differences observed in the potential surfaces of RFP[24], KFP[23] and GFP[9,23,31] chromophore models. It also explains why pyramidalized bridge carbons are observed at twisted intersections in HBDI[23] and HBI[31] – the bridge-charged diabatic state projects on the $S_1$ state in both cases, and methine anion prefers $sp^3$ hybridization.

For constant splitting within the upper block and between the lower and mean upper block diagonals, degeneracy could be achieved by increasing the coupling within the upper block. One way this might be achieved would be by compressing the planar bond. Stretching the same bond, on the other hand, would raise the energy of the $S_0$ state, and may also induce degeneracy. Structures of SA-CASSCF intersections of HBDI tend to suggest the latter mechanism[23]. The branching planes of twisted conical intersections in fluorescent protein chromophores always contain a bridge stretch component and a torsion component[9,24,31]. Stretching models are also important at planar (optically active) geometries. Bridge stretching modes are prominent in the resonance Raman spectrum of HBDI anion,[58,89] and resonance and pre-resonance Raman spectra of GFPs[89,90]. There is spectroscopic evidence of asymmetric vibrations in the spectroscopy of other fluorogenic monomethine dyes[91]. The resonance in Figure 1 suggests that bond stretching will play an important role.

## 10. Limits of Applicability

Our analysis of the photoisomerization ansatz rests on the existence of a particular solution to the state-averaged CASSCF problem for HBDI anion. One way to probe the limits of the ansatz is to investigate the limits of stability and existence of the SA-CASSCF solution itself.

There are two main ways in which the SA-CASSCF solution may break down.

The first type of breakdown will occur if the particular solution becomes unstable in a particular region of configuration space – that is, if the SA-CASSCF variational problem converges to a different solution given an initial guess which is arbitrarily close to the solution of interest. This can be diagnosed by a sudden change of the character of the orbitals, and is related to the non-analyticity of self-consistent field solutions[92]. The state-average energy (the variational objective function) may not identify the most appropriate solution in this case, because there may be excitations that lower the average energy but describe a higher excited state (for example, core excitations may lower both state energies, and the average, but cannot reasonably be assigned to a UV/VIS excitation).

A second type of breakdown will occur if there is a energy crossing between states in the upper and lower block of the block-diagonalized Hamiltonian. This is a breakdown of the diabatic character of the block diagonalization. It will manifest itself in a sharp change in the character of one of the diabatic states.

We have found, generally, that the SA-CASSCF solution in Figure 2 is robust against both types of breakdown. The SA-CASSCF solution depicted in figure 1 can be obtained over the entire domain of bridge torsion. For geometries described here, there is more evidence in the Supplement[55]. Furthermore, when we have observed convergence to different solutions to the SA3-CAS(4,3) problem, the breakdown arises from a lowering of the $S_2$ state that overrides a slight rise of the $S_0$ and $S_1$ energies. From this we infer that even when the diabatic-state picture breaks down for the lowest three states of HBDI, it may still provide a good description of the lowest two states. This is also consistent with our observation that a qualitatively very similar solution can also be found to the *two*-state average problem (SA2-CAS(4,3)). This suggests that the three-state ansatz is nearly always a good model for the $S_1$ and $S_0$ states (and therefore the photoisomerization).

The second type of failure occurs when states in the disjoint blocks of the Hamiltonian cross each other, leading to a breakdown of the diabatic character of the block-diagonalization transform. This does not occur at any of the geometries that we have generated by constrained or unconstrained relaxation on the $S_0$ or $S_1$ states. We have found that it can be induced by rigid twisting of both bridge bonds (in con- or disrotatory fashion) to a near perpendicular conformation while holding all other geometric parameters at their ground state minimum values. This induces a crossing between the $|B\rangle$ state and an ionic state dominated by the $|ppii\rangle$ CSF. This changes the character of the $S_2$ adiabatic state (but not appreciably the $S_1$ state). The ground and excited state energies at these geometries are quite high, so we do not think they are relevant to the photochemical dynamics. We do not observe this behavior if the molecule, held at a similar distribution of bridge torsions, is allowed to relax on the $S_1$ or $S_0$ surface. From this, we infer that breakdown of the diabatic-state picture due to inter-block crossing is unlikely to become relevant in models of the photoisomerization process. We do note, however, that recent experimental evidence suggests this may occur for the *ground-state* isomerization, if it is catalyzed by the presence of a base[93].

When we do observe a breakdown of the SA-CASSCF solution space (as outlined above), there is usually no effect on the $S_1$ or $S_0$ states, but the character of the $S_2$ state changes. Also, breakdowns occur at values of acute torsion, which are not likely to be accessed that in the early photodynamics. Taking a strong interpretation of Kasha's rule[94-96], that $S_n$-$S_1$ relaxation is the fastest process of interest in the system, leads to the conclusion that any intruder states impinging on $S_2$ at these geometries are irrelevant to the photoisomerization. We have not found, and are not aware of, any indication that Kasha's rule is violated in these systems. The similarity of fluorescence induced by one and two-photon excitation in HBDI and GFP variants[97-99] argues in support of the rule.

In conclusion, if we take the breakdown of the SA-CASSCF solution in Figure 1 as an indicator of the limits of applicability of the three-state ansatz described here, we readily infer that the ansatz provides a very robust foundation for models of the bridge photoisomerization process in GFP chromophores. Indeed, since similar SA-CASSCF solutions exist for chromophores from other fluorescent protein subfamilies[23], the ansatz may be considerably more general than indicated here.

## 10. Conclusion

We have described the photoisomerization reaction of a green fluorescent protein chromophore model using a three state-diabatic representation. The diabatic states emerge naturally when a unitary block diagonalization transform is applied to a solution of the variational SA-CASSCF problem. They possess a simple structure and retain a charge-localized valence-bond form over relevant regions of the potential energy surfaces. This structure suggests that parametric Hamiltonians based on valence bond theory may provide a realistic model of the photoisomerization reaction.

The structure of the energy eigenstates in the diabatic representation suggests that a 'twin state' model[74] may be successfully applied to the optical excitation event, but cannot fully describe the photoisomerization. In order to describe *both* photoisomerization pathways, a three-state model is required. This is because photoisomerization in these systems can occur in *either* of the bonds adjoining the methine bridge, or possibly a combination[9,23,25]. A model built from two valence-bond diagrams cannot span sufficient electronic structure to describe photoisomerization via both possible pathways. However a three-state model *can* span this space. We have illustrated this by a simple example of a model parametric Hamiltonian in the three-state representation. Preliminary results suggest an analogous SA-CASSCF solution can be found for several other examples of fluorogenic monomethine dyes[14,100]. We anticipate our approach may also be relevant to these dyes, and will facilitate new understanding of their photodynamics in biological and other condensed matter.

**Acknowledgement** This work was supported by Australian Research Council (ARC) Discovery Project DP0877875. All computations were carried out at the National Computational Infrastructure (NCI) National Facility in Canberra. We thank the NCI staff for their assistance and expertise. Time on the NCI machines was generously provided through a Merit Allocation Scheme (MAS) Grant (Project M03). Some of the figures were generated using VMD[101], Mathematica[102] and ChemBioDraw Ultra[103]. We thank N.S. Hush, A. Jacko, T.J. Martínez, J. Michl, B. Powell, L. Radom, H.F. Schaefer III, S. Shaik and M. Smith, for helpful discussions. We also thank J.R. Reimers and M.J.T. Jordan for many constructive suggestions on a draft manuscript.

**Supplementary Information Available.** Supplementary material, including a description of the torsion surface fitting leading to Figure 6, and raw electronic structure results sufficient for reproduction of the calculations are available via EPAPS[55].

**Tables**

**Table 1.** Comparison of published estimates of the $S_0$-$S_1$ splitting of HBDI and HBI anion obtained with SA-CASSCF and MR-RSPT2 methods, and the peak of the gas-phase absorption spectrum of HBDI. Sources are indicated in the far right column.

| Model | CASSCF | Basis | MRPT2[a] | $\Delta E_{0-1}$(eV) | Ref. #'s |
|---|---|---|---|---|---|
| HBDI | SA3-CAS(4,3) | DZP | MS | 2.69 | 23 |
| HBDI | SA2-CAS(12,11) | 6-31g* | SS | 2.51 | 58 |
| HBDI | SA2-CAS(12,11) | 6-31g* | SS | 2.35 | 60 |
| HBDI | Experiment | – | – | 2.59 | 63 |
| HBI | SA2-CAS(2,2) | 6-31g* | SS | 2.66 | 31 |
| HBI | SA2-CAS(12,11) | 6-31g* | SS | 2.67 | 9, 61 |

[a]SS=Single State, MS=Multi-State

**Table 2.** Density matrix elements of the $|P\rangle$ diabatic state, represented in the basis of configuration state functions (CSFs) over fragment orbitals on the phenoxy ($p$), methine bridge ($b$) and imidazolinone ($i$), at representative geometries of HBDI anion. Geometries include ground state ($S_0$) minima of the Z and E isomers (Z,E-Min-$S_0$), phenoxy-twisted $S_1$ minima of the Z and E isomers (Z,E-P-$S_1$), an imidazolinone-twisted $S_1$ minimum (I-$S_1$) and a constrained, $S_1$-relaxed Hula-Twist geometry (HT-$S_1$). Regardless of geometry, $|P\rangle$ is dominated by the CSF with a doubly occupied phenoxy ($p$) fragment orbital.

| | Z-P-$S_1$ | Z-Min-$S_0$ | I-$S_1$ | E-Min-$S_0$ | E-P-$S_1$ | HT-$S_1$ |
|---|---|---|---|---|---|---|
| $\langle ppbi|P\rangle\langle P|ppbi\rangle$ | 0.70 | 0.68 | 1.00 | 0.69 | 0.70 | 1.00 |
| $\langle bbpi|P\rangle\langle P|bbpi\rangle$ | 0.00 | 0.00 | 0.00 | 0.00 | 0.00 | 0.00 |
| $\langle iipb|P\rangle\langle P|iipb\rangle$ | 0.00 | 0.00 | 0.00 | 0.00 | 0.00 | 0.00 |
| $\langle ppbi|P\rangle\langle P|bbpi\rangle$ | 0.00 | 0.05 | 0.00 | 0.05 | 0.00 | 0.00 |
| $\langle bbpi|P\rangle\langle P|iipb\rangle$ | 0.00 | 0.00 | 0.00 | 0.00 | 0.00 | 0.00 |
| $\langle ppbi|P\rangle\langle P|iipb\rangle$ | 0.00 | -0.05 | 0.00 | 0.05 | 0.00 | 0.00 |
| $\langle bbii|P\rangle\langle P|bbii\rangle$ | 0.00 | 0.01 | 0.00 | 0.01 | 0.00 | 0.00 |
| $\langle ppii|P\rangle\langle P|ppii\rangle$ | 0.20 | 0.18 | 0.00 | 0.18 | 0.20 | 0.00 |
| $\langle ppbb|P\rangle\langle P|ppbb\rangle$ | 0.10 | 0.12 | 0.00 | 0.11 | 0.10 | 0.00 |
| $\langle bbii|P\rangle\langle P|ppii\rangle$ | 0.00 | -0.04 | 0.00 | -0.05 | 0.00 | 0.00 |
| $\langle ppii|P\rangle\langle P|ppbb\rangle$ | 0.14 | 0.15 | 0.00 | 0.14 | 0.14 | 0.00 |
| $\langle ppbi|P\rangle\langle P|ppbi\rangle$ | 0.00 | -0.04 | 0.00 | -0.04 | 0.00 | 0.00 |
| $\langle ppbi|P\rangle\langle P|bbii\rangle$ | 0.00 | 0.09 | 0.00 | -0.09 | 0.00 | 0.00 |
| $\langle ppbi|P\rangle\langle P|ppii\rangle$ | -0.38 | -0.35 | 0.01 | 0.35 | 0.37 | 0.03 |
| $\langle ppbi|P\rangle\langle P|ppbb\rangle$ | -0.26 | -0.29 | 0.01 | 0.28 | 0.27 | 0.01 |
| $\langle bbpi|P\rangle\langle P|bbii\rangle$ | 0.00 | 0.01 | 0.00 | -0.01 | 0.00 | 0.00 |
| $\langle bbpi|P\rangle\langle P|ppii\rangle$ | 0.00 | -0.02 | 0.00 | 0.03 | 0.00 | 0.00 |
| $\langle bbpi|P\rangle\langle P|ppbb\rangle$ | 0.00 | -0.02 | 0.00 | 0.02 | 0.00 | 0.00 |
| $\langle iipb|P\rangle\langle P|bbii\rangle$ | 0.00 | -0.01 | 0.00 | -0.01 | 0.00 | 0.00 |
| $\langle iipb|P\rangle\langle P|ppii\rangle$ | 0.00 | 0.02 | 0.00 | 0.02 | 0.00 | 0.00 |
| $\langle iipb|P\rangle\langle P|ppbb\rangle$ | 0.00 | 0.02 | 0.00 | 0.02 | 0.00 | 0.00 |

**Table 3.** Density matrix elements of the $|B\rangle$ diabatic state, represented in the basis of configuration state functions (CSFs) over the fragment orbitals on the phenoxy ($p$), methine bridge ($b$) and imidazolinone ($i$), at representative geometries of HBDI anion. Geometries include ground state ($S_0$) minima of the Z and E isomers (Z,E-Min-$S_0$), phenoxy-twisted $S_1$ minima of the Z and E isomers (Z,E-P-$S_1$), an imidazolinone-twisted $S_1$ minimum (I-$S_1$) and a constrained, $S_1$-relaxed Hula-Twist geometry (HT-$S_1$). Regardless of geometry, $|B\rangle$ is dominated by the CSF with a doubly occupied methine bridge ($b$) fragment orbital.

| | Z-P-$S_1$ | Z-Min-$S_0$ | I-$S_1$ | E-Min-$S_0$ | E-P-$S_1$ | HT-$S_1$ |
|---|---|---|---|---|---|---|
| $\langle ppbi|B\rangle\langle B|ppbi\rangle$ | 0.00 | 0.00 | 0.00 | 0.00 | 0.00 | 0.00 |
| $\langle bbpi|B\rangle\langle B|bbpi\rangle$ | 1.00 | 0.94 | 1.00 | 0.94 | 1.00 | 1.00 |
| $\langle iipb|B\rangle\langle B|iipb\rangle$ | 0.00 | 0.00 | 0.00 | 0.00 | 0.00 | 0.00 |
| $\langle ppbi|B\rangle\langle B|bbpi\rangle$ | 0.00 | 0.06 | 0.00 | 0.06 | 0.00 | 0.00 |
| $\langle bbpi|B\rangle\langle B|iipb\rangle$ | 0.00 | 0.05 | 0.00 | -0.05 | 0.00 | 0.00 |
| $\langle ppbi|B\rangle\langle B|iipb\rangle$ | 0.00 | 0.00 | 0.00 | 0.00 | 0.00 | 0.00 |
| $\langle bbii|B\rangle\langle B|bbii\rangle$ | 0.00 | 0.00 | 0.00 | 0.00 | 0.00 | 0.00 |
| $\langle ppii|B\rangle\langle B|ppii\rangle$ | 0.00 | 0.05 | 0.00 | 0.05 | 0.00 | 0.00 |
| $\langle ppbb|B\rangle\langle B|ppbb\rangle$ | 0.00 | 0.00 | 0.00 | 0.00 | 0.00 | 0.00 |
| $\langle bbii|B\rangle\langle B|ppii\rangle$ | 0.00 | 0.01 | 0.00 | 0.01 | 0.00 | 0.00 |
| $\langle ppii|B\rangle\langle B|ppbb\rangle$ | 0.00 | 0.00 | 0.00 | 0.01 | 0.00 | 0.00 |
| $\langle ppbi|B\rangle\langle B|ppbi\rangle$ | 0.00 | 0.00 | 0.00 | 0.00 | 0.00 | 0.00 |
| $\langle ppbi|B\rangle\langle B|bbii\rangle$ | 0.00 | 0.00 | 0.00 | 0.00 | 0.00 | 0.00 |
| $\langle ppbi|B\rangle\langle B|ppii\rangle$ | 0.00 | 0.01 | 0.00 | -0.01 | 0.00 | 0.00 |
| $\langle ppbi|B\rangle\langle B|ppbb\rangle$ | 0.00 | 0.00 | 0.00 | 0.00 | 0.00 | 0.00 |
| $\langle bbpi|B\rangle\langle B|bbii\rangle$ | 0.00 | 0.03 | 0.00 | -0.02 | 0.00 | -0.01 |
| $\langle bbpi|B\rangle\langle B|ppii\rangle$ | 0.00 | 0.23 | -0.01 | -0.23 | 0.00 | 0.01 |
| $\langle bbpi|B\rangle\langle B|ppbb\rangle$ | 0.00 | 0.02 | -0.02 | -0.03 | 0.00 | -0.03 |
| $\langle iipb|B\rangle\langle B|bbii\rangle$ | 0.00 | 0.00 | 0.00 | 0.00 | 0.00 | 0.00 |
| $\langle iipb|B\rangle\langle B|ppii\rangle$ | 0.00 | 0.01 | 0.00 | 0.01 | 0.00 | 0.00 |
| $\langle iipb|B\rangle\langle B|ppbb\rangle$ | 0.00 | 0.00 | 0.00 | 0.00 | 0.00 | 0.00 |

**Table 4.** Density matrix elements of the $|I\rangle$ diabatic state, represented in the basis of configuration state functions (CSFs) over the fragment orbitals on the phenoxy ($p$), methine bridge ($b$) and imidazolinone ($i$), at representative geometries of HBD I anion. Geometries include ground state ($S_0$) minima of the Z and E isomers (Z,E-Min-$S_0$), phenoxy-twisted $S_1$ minima of the Z and E isomers (Z,E-P-$S_1$), an imidazolinone-twisted $S_1$ minimum (I-$S_1$) and a constrained, $S_1$-relaxed Hula-Twist geometry (HT-$S_1$). Regardless of geometry, $|I\rangle$ is dominated by the CSF with a doubly occupied imidazolinone ($i$) fragment orbital.

| | Z-P-$S_1$ | Z-Min-$S_0$ | I-$S_1$ | E-Min-$S_0$ | E-P-$S_1$ | HT-$S_1$ |
|---|---|---|---|---|---|---|
| $\langle ppbi|I\rangle\langle I|ppbi\rangle$ | 0.00 | 0.00 | 0.00 | 0.00 | 0.00 | 0.00 |
| $\langle bbpi|I\rangle\langle I|bbpi\rangle$ | 0.00 | 0.00 | 0.00 | 0.00 | 0.00 | 0.00 |
| $\langle iipb|I\rangle\langle I|iipb\rangle$ | 1.00 | 0.71 | 0.70 | 0.71 | 1.00 | 1.00 |
| $\langle ppbi|I\rangle\langle I|bbpi\rangle$ | 0.00 | 0.00 | 0.00 | 0.00 | 0.00 | 0.00 |
| $\langle bbpi|I\rangle\langle I|iipb\rangle$ | 0.00 | 0.05 | 0.00 | -0.05 | 0.00 | 0.00 |
| $\langle ppbi|I\rangle\langle I|iipb\rangle$ | 0.00 | -0.05 | 0.00 | 0.05 | 0.00 | 0.00 |
| $\langle bbii|I\rangle\langle I|bbii\rangle$ | 0.00 | 0.12 | 0.09 | 0.12 | 0.00 | 0.00 |
| $\langle ppii|I\rangle\langle I|ppii\rangle$ | 0.00 | 0.15 | 0.21 | 0.15 | 0.00 | 0.00 |
| $\langle ppbb|I\rangle\langle I|ppbb\rangle$ | 0.00 | 0.01 | 0.00 | 0.01 | 0.00 | 0.00 |
| $\langle bbii|I\rangle\langle I|ppii\rangle$ | 0.00 | 0.14 | 0.13 | 0.14 | 0.00 | 0.00 |
| $\langle ppii|I\rangle\langle I|ppbb\rangle$ | 0.00 | -0.04 | 0.00 | -0.04 | 0.00 | 0.00 |
| $\langle ppbi|I\rangle\langle I|ppbi\rangle$ | 0.00 | -0.04 | 0.00 | -0.04 | 0.00 | 0.00 |
| $\langle ppbi|I\rangle\langle I|bbii\rangle$ | 0.00 | 0.02 | 0.00 | -0.02 | 0.00 | 0.00 |
| $\langle ppbi|I\rangle\langle I|ppii\rangle$ | 0.00 | 0.02 | 0.00 | -0.02 | 0.00 | 0.00 |
| $\langle ppbi|I\rangle\langle I|ppbb\rangle$ | 0.00 | -0.01 | 0.00 | 0.01 | 0.00 | 0.00 |
| $\langle bbpi|I\rangle\langle I|bbii\rangle$ | 0.00 | -0.02 | 0.00 | 0.02 | 0.00 | 0.00 |
| $\langle bbpi|I\rangle\langle I|ppii\rangle$ | 0.00 | -0.02 | 0.00 | 0.02 | 0.00 | 0.00 |
| $\langle bbpi|I\rangle\langle I|ppbb\rangle$ | 0.00 | 0.01 | 0.00 | -0.01 | 0.00 | 0.00 |
| $\langle iipb|I\rangle\langle I|bbii\rangle$ | 0.00 | -0.29 | -0.25 | -0.29 | 0.00 | 0.02 |
| $\langle iipb|I\rangle\langle I|ppii\rangle$ | 0.00 | -0.33 | -0.39 | -0.33 | 0.00 | 0.05 |
| $\langle iipb|I\rangle\langle I|ppbb\rangle$ | 0.00 | 0.09 | 0.00 | 0.09 | 0.00 | 0.00 |

**Table 5.** Density matrix elements of the lowest adiabatic state $S_0$ at representative geometries of HBDI anion. Geometries include ground state ($S_0$) minima of the $Z$ and $E$ isomers ($Z,E$-Min-$S_0$), phenoxy-twisted $S_1$ minima of the $Z$ and $E$ isomers ($Z,E$-P-$S_1$), an imidazolinone-twisted $S_1$ minimum (I-$S_1$) and a constrained, $S_1$-relaxed Hula-Twist geometry (HT-$S_1$).

| Geo. | $Z$-P-$S_1$ | $Z$-Min-$S_0$ | I-$S_1$ | $E$-Min-$S_0$ | $E$-P-$S_1$ | HT-$S_1$ |
|---|---|---|---|---|---|---|
| $\langle P|S_0\rangle\langle S_0|P\rangle$ | 1.00 | 0.47 | 0.00 | 0.45 | 1.00 | 0.50 |
| $\langle B|S_0\rangle\langle S_0|B\rangle$ | 0.00 | 0.07 | 0.00 | 0.07 | 0.00 | 0.00 |
| $\langle I|S_0\rangle\langle S_0|I\rangle$ | 0.00 | 0.46 | 1.00 | 0.48 | 0.00 | 0.50 |
| $\langle P|S_0\rangle\langle S_0|B\rangle$ | 0.00 | 0.18 | 0.00 | 0.17 | 0.00 | 0.02 |
| $\langle B|S_0\rangle\langle S_0|I\rangle$ | 0.00 | 0.18 | 0.00 | -0.18 | 0.00 | -0.02 |
| $\langle P|S_0\rangle\langle S_0|I\rangle$ | 0.00 | 0.46 | -0.01 | -0.47 | 0.00 | -0.50 |

**Table 6.** Density matrix elements of the first excited adiabatic state $S_1$ at representative geometries of HBDI anion. Geometries include ground state ($S_0$) minima of the Z and E isomers (Z,E-Min-$S_0$), phenoxy-twisted $S_1$ minima of the Z and E isomers (Z,E-P-$S_1$), an imidazolinone-twisted $S_1$ minimum (I-$S_1$) and a constrained, $S_1$-relaxed Hula-Twist geometry (HT-$S_1$).

| Geo. | Z-P-$S_1$ | Z-Min-$S_0$ | I-$S_1$ | E-Min-$S_0$ | E-P-$S_1$ | HT-$S_1$ |
|---|---|---|---|---|---|---|
| $\langle P\|S_1\rangle\langle S_1\|P\rangle$ | 0.00 | 0.52 | 0.66 | 0.55 | 0.00 | 0.50 |
| $\langle B\|S_1\rangle\langle S_1\|B\rangle$ | 0.39 | 0.03 | 0.34 | 0.03 | 0.37 | 0.00 |
| $\langle I\|S_1\rangle\langle S_1\|I\rangle$ | 0.61 | 0.45 | 0.00 | 0.43 | 0.63 | 0.50 |
| $\langle P\|S_1\rangle\langle S_1\|B\rangle$ | 0.00 | -0.12 | 0.47 | -0.12 | 0.00 | 0.02 |
| $\langle B\|S_1\rangle\langle S_1\|I\rangle$ | 0.49 | 0.11 | 0.01 | -0.11 | -0.48 | 0.02 |
| $\langle P\|S_1\rangle\langle S_1\|I\rangle$ | 0.00 | -0.48 | 0.01 | 0.48 | 0.00 | 0.50 |

**Table 7.** Density matrix elements of the second excited adiabatic state $S_2$ at representative geometries of HBDI anion. Geometries include ground state ($S_0$) minima of the Z and E isomers (Z,E-Min-$S_0$), phenoxy-twisted $S_1$ minima of the Z and E isomers (Z,E-P-$S_1$), an imidazolinone-twisted $S_1$ minimum (I-$S_1$) and a constrained, $S_1$-relaxed Hula-Twist geometry (HT-$S_1$).

| Geo. | Z-P-$S_1$ | Z-Min-$S_0$ | I-$S_1$ | E-Min-$S_0$ | E-P-$S_1$ | HT-$S_1$ |
|---|---|---|---|---|---|---|
| $\langle P|S_2\rangle\langle S_2|P\rangle$ | 0.00 | 0.00 | 0.34 | 0.00 | 0.00 | 0.00 |
| $\langle B|S_2\rangle\langle S_2|B\rangle$ | 0.61 | 0.90 | 0.66 | 0.91 | 0.63 | 1.00 |
| $\langle I|S_2\rangle\langle S_2|I\rangle$ | 0.39 | 0.10 | 0.00 | 0.09 | 0.37 | 0.00 |
| $\langle P|S_2\rangle\langle S_2|B\rangle$ | 0.00 | -0.06 | -0.47 | -0.05 | 0.00 | -0.04 |
| $\langle B|S_2\rangle\langle S_2|I\rangle$ | -0.49 | -0.29 | 0.00 | 0.29 | 0.48 | 0.01 |
| $\langle P|S_2\rangle\langle S_2|I\rangle$ | 0.00 | 0.02 | 0.00 | -0.02 | 0.00 | 0.00 |

**Table 8.** Elements of the lower block effective Hamiltonian at representative geometries of HBDI anion. Geometries include ground state ($S_0$) minima of the $Z$ and $E$ isomers ($Z,E$-Min-$S_0$), phenoxy-twisted $S_1$ minima of the $Z$ and $E$ isomers ($Z,E$-P-$S_1$), an imidazolinone-twisted $S_1$ minimum (I-$S_1$) and a constrained, $S_1$-relaxed Hula-Twist geometry (HT-$S_1$). All energies are in kcal/mol and are referenced to the mean along the diagonal evaluated at the HT-$S_1$ geometry.

| Geo. | Z-P-$S_1$ | Z-Min-$S_0$ | I-$S_1$ | E-Min-$S_0$ | E-P-$S_1$ | HT-$S_1$ |
|---|---|---|---|---|---|---|
| $\langle P|H_{eff}|P\rangle$ | -62.7 | -54.6 | -17.4 | -50.0 | -58.8 | -9.3 |
| $\langle B|H_{eff}|B\rangle$ | 12.5 | -6.0 | -0.3 | 0.1 | 16.2 | 18.6 |
| $\langle I|H_{eff}|I\rangle$ | -2.1 | -51.2 | -52.7 | -49.5 | -2.3 | -9.3 |
| $\langle P|H_{eff}|B\rangle$ | 0.0 | -13.1 | -25.6 | -12.2 | 0.0 | -1.1 |
| $\langle B|H_{eff}|I\rangle$ | -33.4 | -19.0 | -0.1 | 19.4 | 33.8 | 0.2 |
| $\langle P|H_{eff}|I\rangle$ | 0.0 | -28.2 | 0.3 | 28.4 | 0.0 | 0.9 |

**Figure Captions**

**Figure 1.** The charge-transfer resonance of HBDI anion at two isomeric (Z,E) geometries. The resonance superposes structures wherein location of a formal anionic charge is correlated with double bond alternation.

**Figure 2.** The electronic structure solution that motivates the ansatz. There is a solution to the three-state averaged four-electron-in-three-orbital variational problem which, at convergence, yields an active space spanned by localized fragment orbitals (top) on the phenoxy, methine-bridge and imidazolinone fragments. Over this orbital space are built six singlet configuration state functions (CSFs, bottom). The 'covalent configurations, each of which supports one doubly occupied orbital and one singlet pair, correlate location of the charge with bond alternation (as in the resonance shown in figure 1). The ionic configurations can be created by polarizing the singlet pairs of the covalent configurations. Double headed arrows are used to highlight the relationship. Note the analogy to carbogenic orbitals and valence-bond structures of an allyl anion.

**Figure 3.** The electronic structure of the diabatic states of HBI anion at representative geometries: the $S_0$ minimum of the Z isomer (Z-Min-$S_0$), the phenoxy-twisted $S_1$ minimum of the Z isomer (Z-P-$S_1$), an imidazolinone-twisted $S_1$ minimum (I-$S_1$), and a distotatory (hula) twisted structure relaxed on $S_1$ under bridge torsion constraints (HT-$S_1$). At each geometry, the one-electron basis has been localized using the Boys procedure (left). The Hamiltonian is block-diagonalized using a unitary transformation to yield three diabatic states, whose density matrices in the localized-orbital CSF basis are shown (matrix element labels at bottom). Diabatic states labeled $|P\rangle$, $|B\rangle$ and $|I\rangle$ are dominated by CSFs with double occupation of the phenoxy (p), bridge (b) and imidazolinone (i) orbitals. The figure shows that the diabatic representation is transferrable across geometries with substantially different bridge torsion.

**Figure 4.** Inferring appropriate Lewis structures for the diabatic states, from the structure of their density matrices in the configuration state basis. (Top) At a planar geometry, using as an example the ground state geometry of the Z isomer (Z-Min-$S_0$). At planar geometries, both the |P⟩ and |I⟩ diabats contain ionic contributions indicative of chemical bonding. The |B⟩ state contains substantially less ionic contribution, which contraindicates the presence of a chemical bond. (Bottom) At a twisted geometry, using as an example the imidazolinone-twisted excited state minimum (I-$S_1$). At twisted geometries, the states with singlet pairing across the twisted bond (|P⟩ and |B⟩ in this case) contain virtually no ionic contribution, indicating a diradical structure. The state with singlet pairing across a planar bond (|I⟩ in this case) maintains ionicity appropriate for a chemical bond.

**Figure 5.** Application of the "twin state" model to the planar and twisted configurations of HBDI, using the ground state minimum of the Z isomer (Z-Min-$S_0$) and the imidazolinone-twisted $S_1$ minimum (I-$S_1$) as examples. The twin state picture is usually invoked for two-state systems, such as the resonating Kekulé structures of benzene. The model can be applied in its normal sense to describe the first excitation of HBDI, but cannot be applied within the same set of structures to the geometries appropriate to an excited-state diradical. At these configurations, the $S_0$ and $S_1$ states are no longer "twins", but the $S_1$ and $S_2$ states are.

**Figure 6.** Potential energy surfaces of the three lowest singlet states yielded by fitting a parametric functional form for the effective Hamiltonian to a collection of geometries obtained by minimization on the first excited adiabatic state ($S_1$) under constraint of the bridge torsion angles. The resulting surfaces (top) show favorable twisting of both bonds on the $S_1$ surface. Furthermore, the charge localization that accompanies excited-state twisting is described by the distribution of diabatic populations in the $S_1$ state. This is possible because the diabatic states have charge-localized character which is maintained over the surface. Details of the fitting procedure are described in the Supplement.

Figure 1

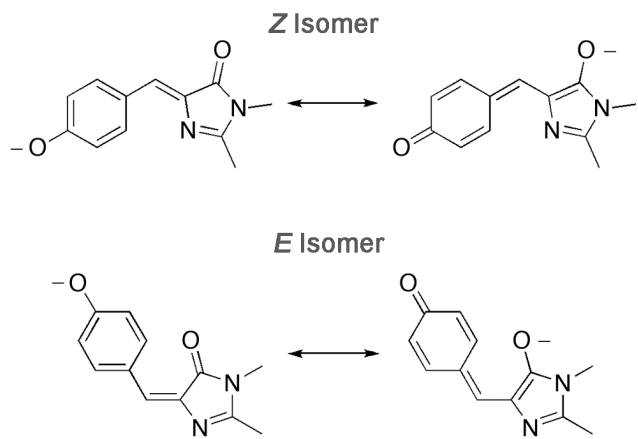

Figure 2

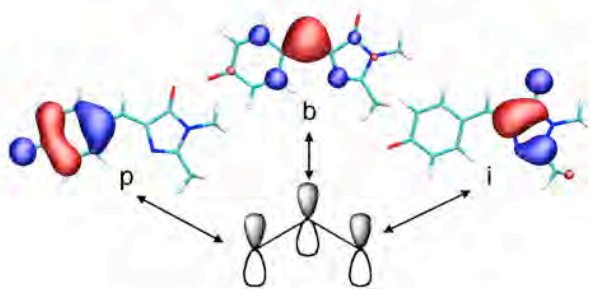
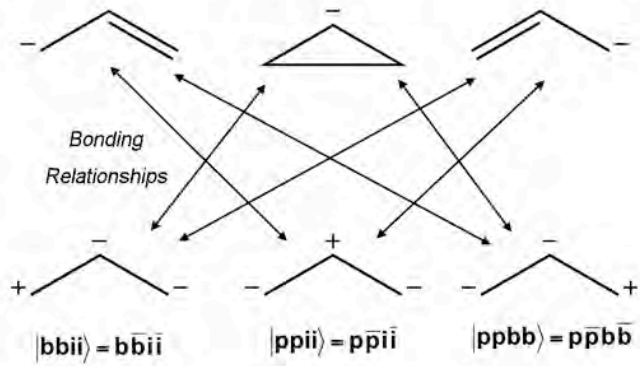

Figure 3

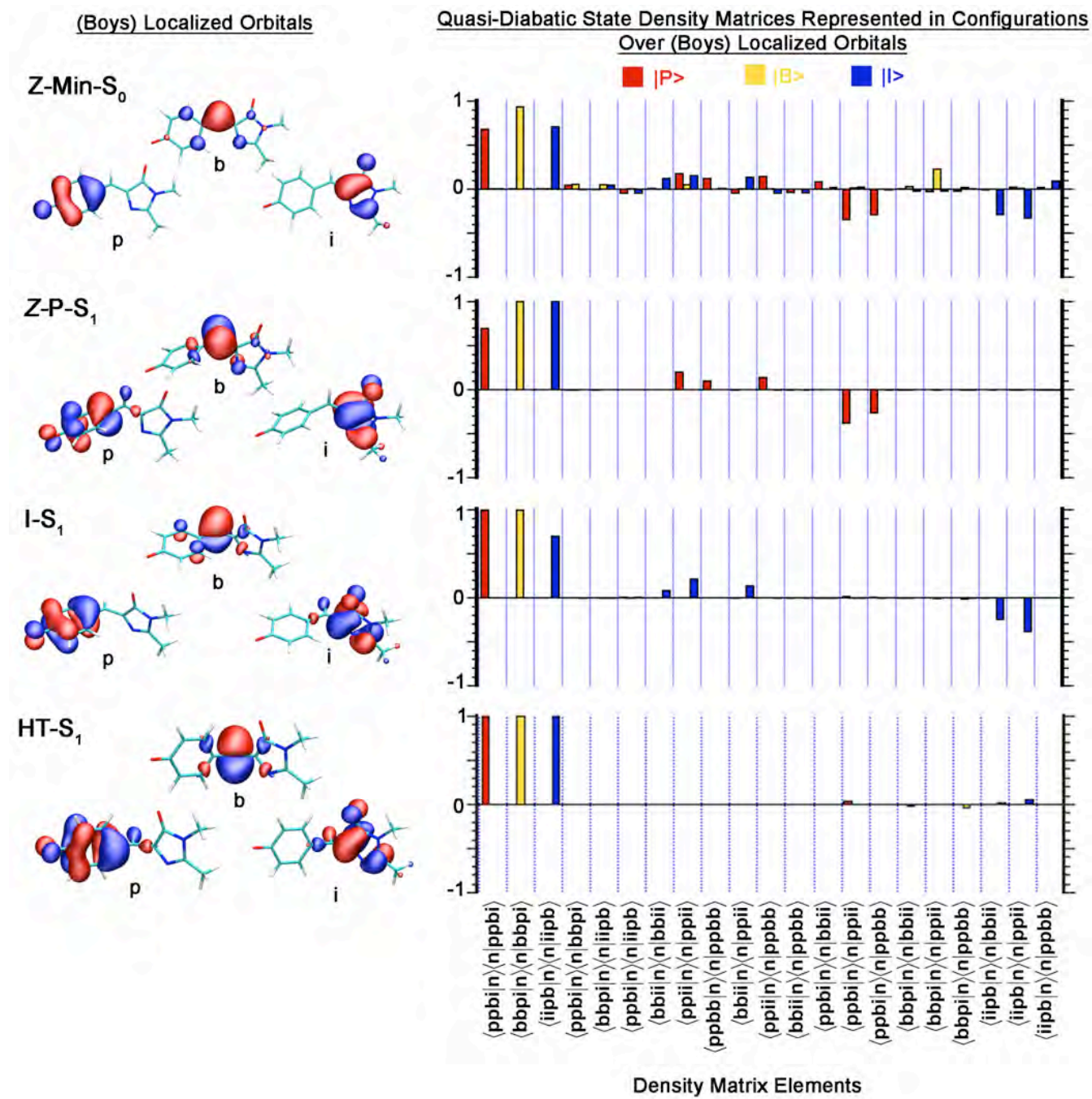

Figure 4

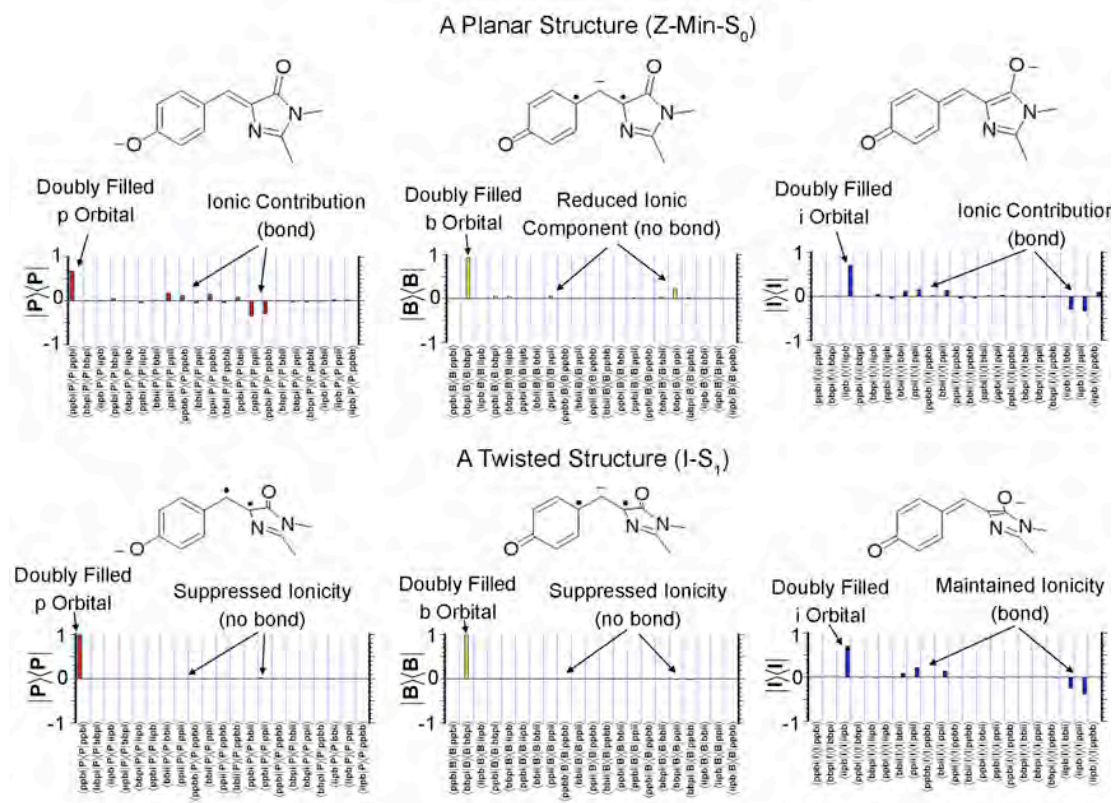

Figure 5

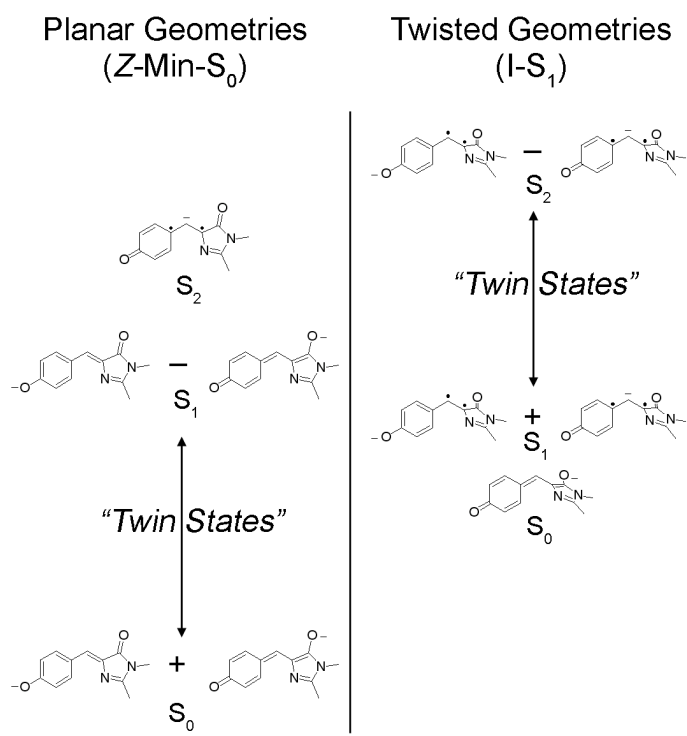

Figure 6

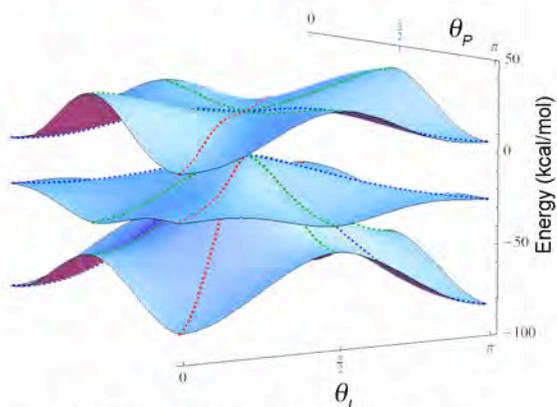

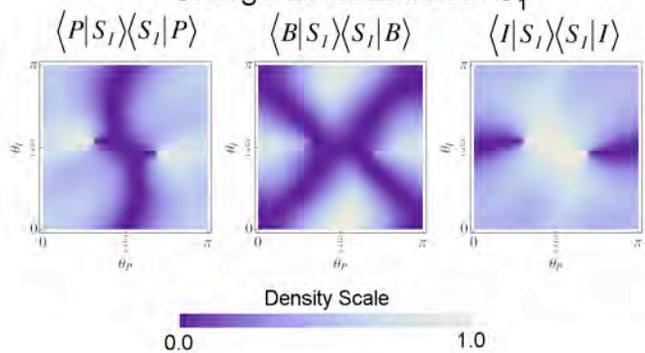